\documentstyle[preprint,aps,prl,epsfig]{revtex}

\newcounter{listfig}

\begin{document}


\title{Universal conductance fluctuations in epitaxial GaMnAs ferromagnets: structural and spin disorder}

\author{L.~Vila$^1$, R.~Giraud$^1$, L.~Thevenard$^1$, A.~Lema\^itre$^1$, F.~Pierre$^1$, J.~Dufouleur$^1$, D.~Mailly$^1$, B.~Barbara$^2$, and G.~Faini$^1$}

\address{$^1$ Laboratoire de Photonique et de Nanostructures - CNRS, Route de Nozay, 91460 Marcoussis, France\\ $^2$ Laboratoire de Magn\'etisme Louis N\'eel - CNRS, BP166, 38042
Grenoble Cedex-09, France}

\date{\today}

\maketitle

\begin{abstract}

\vspace{-3mm}

Mesoscopic transport measurements reveal a large effective phase coherence length in epitaxial GaMnAs ferromagnets, contrary to usual 3$d$-metal ferromagnets. Universal conductance fluctuations  of single nanowires are compared for epilayers with a tailored anisotropy. At large magnetic fields, quantum interferences are due to \emph{structural disorder} only, and an unusual behavior related to hole-induced ferromagnetism is evidenced, for both quantum interferences and decoherence. 
At small fields, phase coherence is shown to persist down to zero field, even in presence of magnons, and an additional \emph{spin disorder} contribution to quantum interferences is observed under domain walls nucleation. 

\vspace{3mm}

\end{abstract}


\narrowtext

Quantum corrections to the conductance were deeply investigated in non magnetic metals for two decades \cite{Imry97,Akkermans05}. It is now well-known that quantum interferences of free carriers in a weakly disordered metallic nanostructure affect the transport properties of a mesoscopic conductor, e.g. giving rise to universal conductance fluctuations (UCF) under an applied magnetic field or to the weak localization of carriers at low fields. In particular, these quantum corrections were used to address decoherence phenomena in a Fermi sea at sub-kelvin temperatures, which are dominated by Coulomb interactions in pure non magnetic metals. Yet, the role of spin was mainly limited to two specific cases: firstly, the weak localization correction becomes an anti-localization when a strong spin-orbit coupling is at work; secondly, efficient spin-flip scattering, even due to minute concentrations of paramagnetic impurities, can strongly alter phase coherence \cite{Benoit88,Benoit92}, leading to an apparent saturation of the decoherence time at very low temperatures (see \cite{Pierre03} and refs. therein). To avoid such paramagnetic fluctuations, which shrink down the inelastic mean free path (and, therefore, the phase coherence length), one can investigate the opposite regime of a much higher magnetic content, when ferromagnetic exchange interactions are large enough to freeze single spin fluctuations. However, quantum transport in a ferromagnet can still be hampered by very efficient decoherence mechanisms, mainly due to low-energy collective spin fluctuations (magnons) and to locally inhomogeneous magnetization states (domain walls).

Recently, a few theoretical \cite{Hong95,Tatara97,Lyanda98,Dugaev01,Tatara01,Tatara03,Adam06} and experimental \cite{Aprili97,Kasai03,Lee04,Wei06} reports investigated mesoscopic transport properties of itinerant ferromagnets. Although theory predicted a specific behavior, due to the interplay between exchange and spin-orbit interactions, for either weak localization \cite{Dugaev01}, Aharonov-Bohm oscillations \cite{Tatara01,Tatara03} or anisotropic UCF \cite{Adam06}, the crucial issue of how to preserve phase coherence in a ferromagnet was often put aside. Yet, only very short phase coherence lengths $L_{\phi}$ were reported from experiments for different 3$d$-metal ferromagnets, $L_{\phi}$ being usually smaller than 30~nm at $T=30$~mK \cite{Aprili97,Kasai03,Lee04,Wei06}. In these granular materials, only a small quantum correction to the conductance was observed, together with a much larger classical contribution of the anisotropic magneto-conductance. Indeed, weak decoherence in a ferromagnet should require both the freezing of spin waves excitations and a saturated magnetization at grain boudaries, though their relative role on decoherence is not known. In the opposite regime of paramagnetism, a short phase coherence length was also found in a diluted magnetic semiconductor (DMS) \cite{Jaro95} due to a strong decoherence induced by single-spin fluctuations.  

In this work, we demonstrate that both the crystalline quality and the anisotropy play a significant role on quantum interferences in a ferromagnet. From UCF measurements, we evidence the unusually large phase coherence length in an epitaxial ferromagnet, GaMnAs \cite{Ohno99}, with $L_{\phi}\approx$~100~nm at $T=100$~mK, which is preserved down to zero field. 
Based on a comparison between epilayers having either an in-plane or a perpendicular magnetic anisotropy, the contributions from \emph{both structural and spin disorder} to UCF can be separated. At large applied fields, quantum interferences are only due to structural disorder and show a behavior specific to hole-induced ferromagnetism in a DMS, quite similarly to a recent study of ultra-narrow GaMnAs nanowires \cite{Wagner06}. Besides, we found that the decoherence mechanism has no apparent dimensional crossover and seems to be independent of the anisotropy. At small fields and for an in-plane anisotropy, the nucleation of rare domain walls clearly affect the correlation field of UCF, although their amplitude is not modified. This strongly suggests that such a weak spin disorder contributes to dephasing without increasing decoherence significantly, even in presence of magnons. 

High quality GaMnAs epilayers, 50~nm thick and with a Mn content of about 6\%, were grown at low temperature ($T\approx$~250~$^{\circ}$C) over either an undoped GaAs buffer layer (compressive strains, in-plane anisotropy) or an undoped InGaAs one (tensile strains, perpendicular anisotropy), previously elaborated at high temperature (see ref.~\cite{Thevenard06} for details). Highly $p$-doped samples were obtained after annealing at $T\approx$~250~$^{\circ}$C for an hour, the Curie temperature thus rising from about 70~K to 130~K. A hole density $p\approx$~5.$10^{20}$~cm$^{-3}$ was deduced from high-field Hall measurements, together with a typical resistivity $\rho \approx$~2~m$\Omega$.cm at $T=$~100~mK for annealed samples ($R_{\Box}\approx$~500~$\Omega$). Despite a much lower carrier density, the mobility is quite comparable to a metal, and a rough estimation of an average value of the diffusion constant gives $D\approx$~10$^{-4}$~m$^2$.s$^{-1}$. 
Narrow nanowires were patterned by ebeam-lithography and Ar-plasma etching, with a length and a width down to 200~nm and 50~nm respectively. The smallest dimension remains longer than the screening length to avoid complications due to edge depletion (within the 10~nm range). Ohmic contacts were made by Ti/Au deposition.
Low-noise four-probe transport measurements were performed down to the base temperature $T_{cryo}\approx25$~mK of a dilution refrigerator, using a lock-in technique and a compensating loop to improve signal to noise ratio, stability and resolution. A magnetic induction up to 7~T was applied perpendicularly to the sample plane. 
All measurements were done with a low-enough voltage bias ($V$) to avoid any increase in the electronic temperature (e$V\le$~k$_BT$) \cite{note1}. 

\begin{figure}
\centerline{\epsfxsize= 15 cm \epsfbox{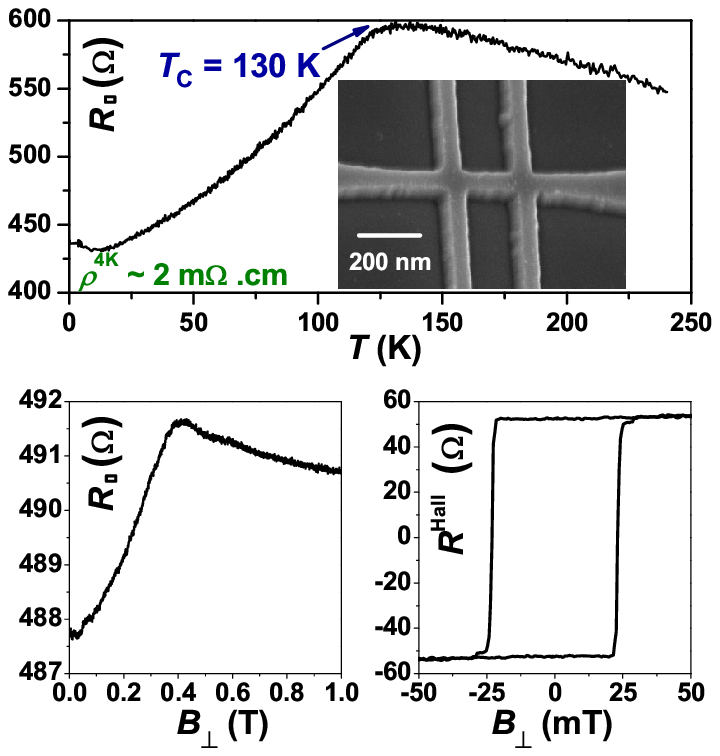}}
\caption{L.~Vila}
\label{fig1}
\end{figure}

For a highly degenerate GaMnAs nanowire, the temperature dependence of the resistivity shows a metallic-like behavior below the Curie temperature (cusp at $T_C\approx$~130~K, see Fig.~\ref{fig1} -top-), with an increase below about 10~K probably due to Coulomb interactions \cite{note1}. The only difference between our two GaMnAs epilayers grown on either GaAs (in-plane anisotropy) or InGaAs (perpendicular anisotropy) is related to the evolution of the magnetization under a perpendicular applied field. This results in a different classical contribution to the magneto conductance below 1~T, as measured with a large applied bias (e$V$/k$_BT\approx$~100 for $I=$~100~nA, at $T=$~100mK). 
For an in-plane anisotropy, a positive anisotropic magneto-resistance (AMR) is observed when the magnetization is continuously rotated out of the plane up to the anisotropy induction $B_A\approx$~0.5~T (see Fig.~\ref{fig1} -bottom left-). For a perpendicular anisotropy, the low-field AMR response does not exist anymore, since the magnetization direction is always perpendicular to the current flow, whereas a strong extraordinary Hall voltage is measured (see Fig.~\ref{fig1} -bottom right-), with an abrupt jump at the coercive induction of about 25~mT. The latter is controlled by the nucleation field of a domain wall, which is typical of a magnetization reversal with a small magnetic viscosity (low density of pinning centers). The strong uniaxial anisotropy also opens a gap of about 3~K in the magnon dispersion curve, so that all spin excitations are frozen at 25~mK. 
In both cases, the high-field negative AMR contribution to the sheet resistance usually becomes smaller than quantum corrections in the mesoscopic regime, contrary to the case of 3$d$ metals, as a consequence of a much larger phase coherence length (see Fig.~\ref{fig3} -top-).

\begin{figure}
\centerline{\epsfxsize= 15 cm \epsfbox{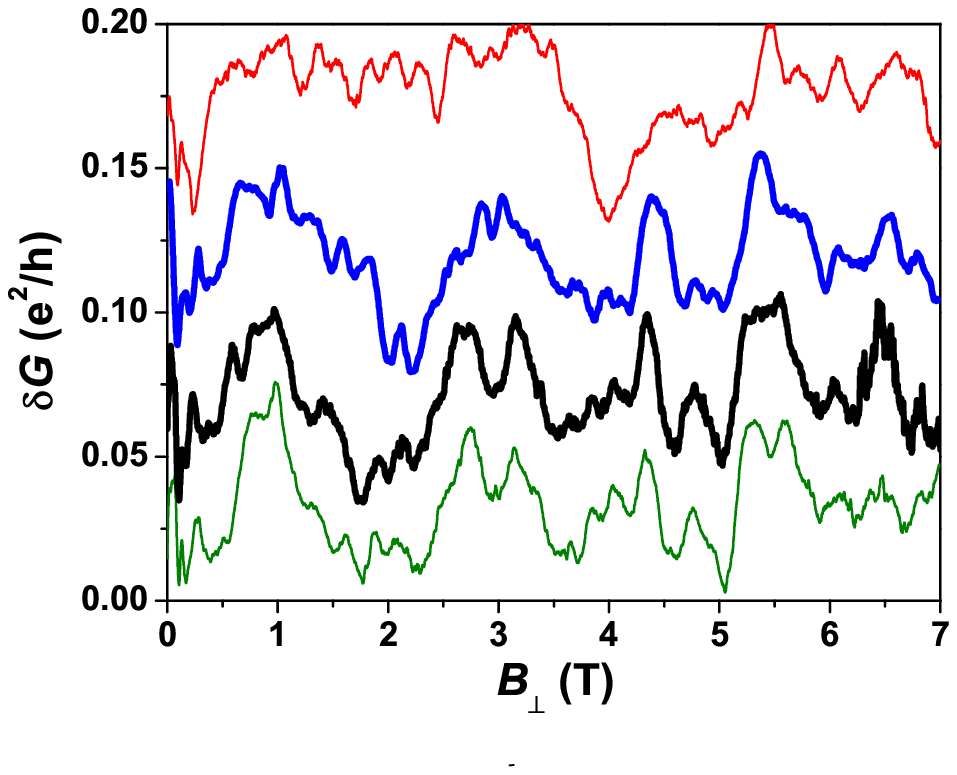}}
\caption{L.~Vila}
\label{fig2}
\end{figure}

\begin{figure}
\centerline{\epsfxsize= 15 cm \epsfbox{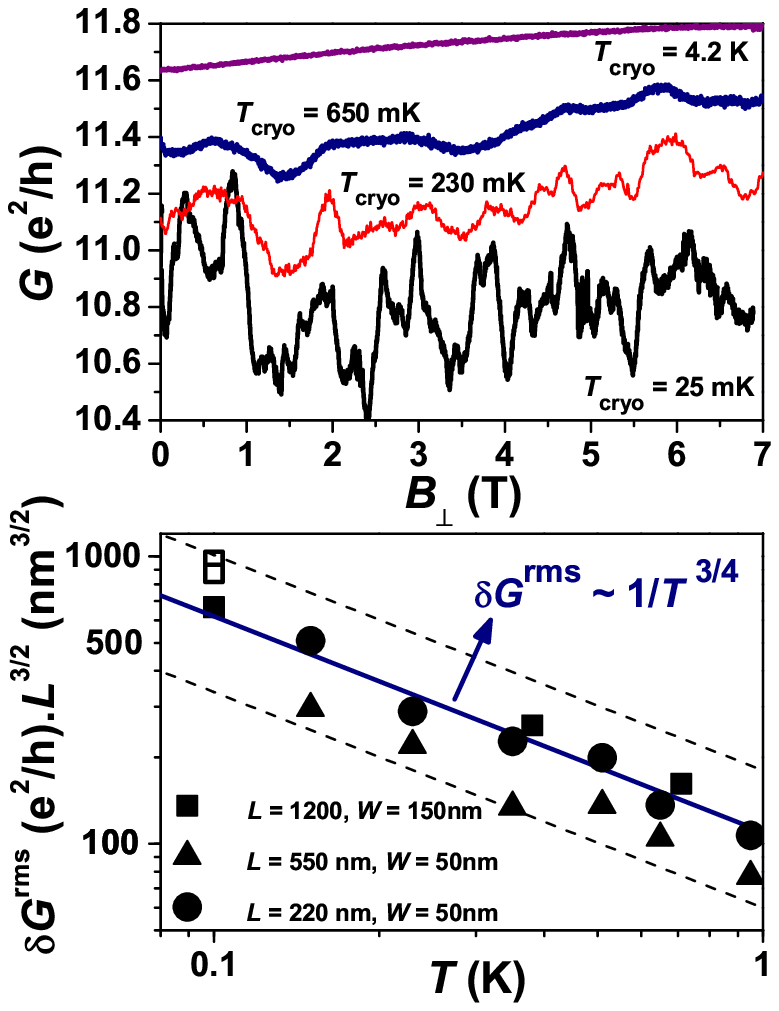}}
\caption{L.~Vila}
\label{fig3}
\end{figure}

We first focus on GaMnAs nanowires with a perpendicular anisotropy, so that the magnetization remains fully saturated down to the remnant state. At $T_{cryo}\approx25$~mK and for a micron-long nanowire, highly reproducible conductance fluctuations are observed when sweeping the magnetic induction between 0 and 7~T, back and forth. 
Fig.~\ref{fig2} shows such four scans, each one being recorded in about 12 hours. The measurement can be reproduced at will (two lower curves), until the temperature is risen (two upper curves). 
A small change in the UCF fingerprint is already observed after a thermal cycling up to $T=$~4.2~K (top thick line), which indicates a small evolution of microscopic disorder. A much stronger effect is observed after a thermal cycling up to room temperature (top thin line). As expected for UCF, similarly to non magnetic metals, a new frozen configuration of structural disorder results in a complete change of the magneto fingerprint, the mean amplitude and the correlation field remaining unchanged. 

The conductance fluctuations were further investigated at higher temperatures (see Fig.~\ref{fig3} -top-) or larger bias, both resulting in a reduction of the amplitude, as expected for UCF. 
Fig.~\ref{fig3} -bottom- shows the scaling behavior of the root mean square deviation to the average conductance, independent of the nanowire length $L$ after renormalization by $L^{3/2}$, including three different lengths with $L\ge L_{\phi}$. No saturation was observed down to our lowest electronic temperature $T_{el}=$~100~mK. Surprisingly, the power-law temperature dependence of the scaling function, with $\delta G^{rms}.L^{3/2}\propto 1/T^{\alpha}$ and $\alpha=0.75\pm0.15$, does not exhibit a dimensional crossover as a function of temperature or of the wire width, for both widths $W=$~50~nm and $W=$~150~nm. 
As the expected thermal length $L_T$ is about 200~nm at $T=$~100~mK and 60~nm at $T=$~1~K, such a crossover from a quasi-1D regime ($L_T\ge W$) to a quasi-2D one ($L_T\le W$) should occur for the larger width $W=$~150~nm, contrary to our observations. Recently, a very similar result on decoherence was obtained by another group \cite{Wagner06} on ultra-narrow nanowires ($W=$~20~nm) having a planar anisotropy, with an identical power law exponent \emph{but} with a smaller amplitude of UCF. This can be expected from edge roughness in narrow nanowires, especially when their width gets closer to the screening length, as in Ref.~\cite{Wagner06}. In our work, the amplitude yields a three times larger phase coherence length, with $L_{\phi}\approx$~100~nm at $T=100$~mK, for both in-plane and perpendicular anisotropy. Such a large value is corroborated by measurements on shorter wires, showing a progressively larger amplitude of UCF, which already gets close to the quantum of conductance for a nanowire length $L=$~220~nm (see Fig.~\ref{fig3}-top-), that is, when $L$ becomes comparable to $L_{\phi}$.  
Note that in Fig.~\ref{fig3}-bottom-, only some \emph{effective} values of $L_{\phi}$ are extracted from the amplitude of UCF using semi-classical results for a quasi-1D behavior \cite{Stone87}, with min($L_T$,$L_{\phi}$)~$\ge W$. Similar values are obtained in two opposite regimes, using either $\delta G^{rms} = e^2/h\cdot (L_\phi/L)^{3/2}$ (for $L_T\ge L_{\phi}$) or $\delta G^{rms} = e^2/h\cdot L_T/L\cdot (L_\phi/L)^{1/2}$ (for $L_T\le L_{\phi}$). This is not a surprise since $L_T\approx L_{\phi}$ for GaMnAs, which further complicates the analysis of the UCF amplitude, even in the framework of the semi-classical approximation. Both results give a $1/\sqrt{T}$ temperature dependence for $L_{\phi}$. Note that, being one to two orders of magnitude lower than in a metal, the diffusion constant is an important limitation to even longer coherence lengths in ferromagnetic DMS. Besides, given that $L_T$ is not so accurately known, the exact nature of the decoherence mechanism remains questionable in a ferromagnetic DMS. 

\begin{figure}
\centerline{\epsfxsize= 15 cm \epsfbox{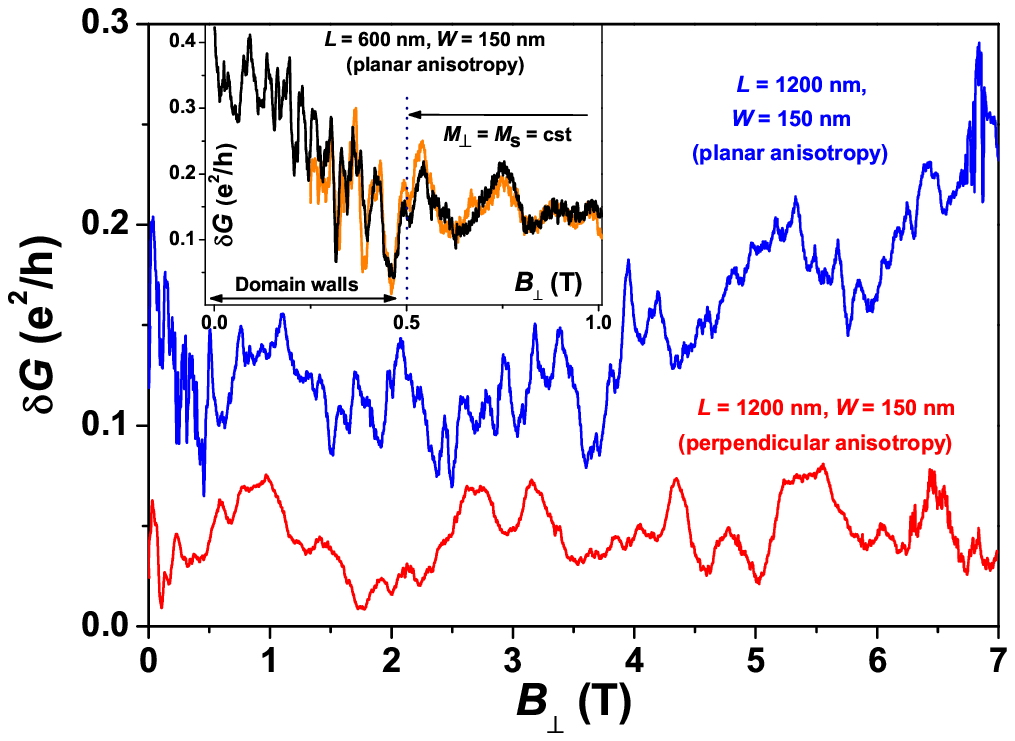}}
\caption{L.~Vila}
\label{fig4}
\end{figure}

Deeper insights on the \emph{nature} of UCF in GaMnAs nanowires were obtained from a detailed comparison between epilayers with either an in-plane or a perpendicular anisotropy (see Fig.~\ref{fig4}). Whereas UCF remain the same down to zero field for the magnet with a strong perpendicular anisotropy (constant and uniform magnetization), a clear crossover is observed below the anisotropy field for an in-plane anisotropy (see Fig.~\ref{fig4} inset). Indeed, when the magnetization rotates back to the sample plane, much faster conductance fluctuations are observed. This results from the influence of magnetic domain walls (spin disorder) on UCF. Just below the anisotropy field, a few domain walls are very likely to spontaneously nucleate on rare strong pinning centers, giving a \emph{reproducible spin contribution} to UCF (see the inset in Fig.~\ref{fig4}, with two successive sweeps, starting from 7~T). Remarkably, the amplitude of conductance fluctuations remains the same, which strongly suggests that the phase coherence length is not affected by the presence of a couple of domain walls. Importantly, the amplitude does not depend on the anisotropy, showing that magnons play a negligible role on decoherence. 
For smaller fields, nucleation of domain walls on weaker pinning centers and depinning result in a hysteretic behavior. Although the amplitude is unchanged, unreproducible fluctuations are observed when measuring minor loops. 
 
Interestingly, the comparison between different anisotropies further reveals some striking deviations from the standard UCF behavior known in non magnetic metals, even when only structural disorder remains at large fields (perpendicular magnetization and strongly reduced spin flip scattering, in any case). For instance, there is not a one-to-one relationship between the amplitude of UCF and a given correlation field.  
Although not understood in detail at present, such a specific behavior in GaMnAs epilayers can be expected for hole-induced ferromagnetism: i) up to four valence subbands can contribute to hole transport, with different spin and band-dependent diffusion coefficients. This may explain the existence of different correlation fields, which can even change with the anisotropy (the latter is fixed by epitaxial strains, which also determine the relative position of valence subbands); ii) the semi-classical approximation may fail since $k_F$.$l$ is only a few units, even in the strongly $p$-doped regime. This makes the exact determination of the phase coherence length even more tedious, and only some effective values are derived from a semi-classical analysis of the UCF amplitude. Note that for ferromagnetic DMS, the Yoffe-Regel criterion does not apply as in conventional metals. The Fermi wavelength is rather large and comparable to the mean free path $l$ (a few nm), giving a small value of $k_F$.$l$ even for highly delocalized carriers.  

To sum up, we evidenced a very large phase coherence length in epitaxial GaMnAs ferromagnets, making a clear distinction between the structural and spin disorder contributions to UCF in this ferromagnet. Effective values of $L_{\phi}$ were extracted from the amplitude of UCF, using semi-classical results, with an additional complication due to different valence subbands contributions.
Furthermore, a behavior specific to hole-induced ferromagnetism was observed, making an accurate analysis of correlation fields non trivial. The exact nature of an unusual decoherence mechanism also remains unclear, with no apparent dimensional crossover for nanowires with a width as large as 150~nm. 
From a comparison between similar epilayers only differing by their anisotropy, the role of spin disorder on UCF was evidenced at small applied fields when domain walls nucleation occurs. 
Below the anisotropy field of a GaMnAs ferromagnet having an in-plane anisotropy, the correlation field of UCF is strikingly modified in presence of a couple of pinned domain walls, whereas their amplitude is unchanged. 
Importantly, no significant decoherence from domain walls or magnons was found in GaMnAs (weak spin disorder regime), although domain walls clearly affect dephasing. 

O.~Mauguin and L.~Largeau are acknowledged for X-ray characterizations of our (Ga,Mn)As epilayers, A.~Miard for epitaxial growth, and L.~Leroy and L.~Couraud for their technical assistance. Financial support by the Conseil G\'en\'eral de l'Essonne and by the national project on nanoscience $\#$~NR216 `PARCOUR' are gratefully acknowledged. We thank the french Agence Nationale pour la Recherche for future financial support of fundamental research.

 \newpage

\centerline{LIST OF FIGURES CAPTIONS}

\begin{list}{FIG.~\arabic{listfig}}{\usecounter{listfig}}

\item (Top) Temperature dependence of the square resistance measured with a GaMnAs nano-Hall bar ($L=$~220~nm, $W=$~50~nm), a common behavior for both anisotropy. (Bottom left) Anisotropic magneto-resistance at $T_{cryo}=$~25~mK, for an in-plane anisotropy. (Bottom right) Extraordinary Hall resistance at $T_{cryo}=$~25~mK, for a perpendicular anisotropy. All measurements were done with a large current $I=$~100~nA, leading to vanishingly small quantum corrections (see text).

\item (Bottom thick line) UCF of a nanowire with a perpendicular anisotropy ($L=$~1200~nm, $W=$~150~nm) measured at base temperature $T_{cryo}\approx25$~mK, with a small current $I=$~1~nA correponding to an electronic temperature $T_{el}\approx$~100~mK \cite{note1}. (Bottom thin line) Same measurement, 24h later. (Top thick line) Same measurement, after a thermal cycling up to $T=$~4.2~K. (Top thin line) Same measurement, after a thermal cycling up to $T=$~300~K.

\item (Top) Temperature dependence of UCF for a short nanowire with a perpendicular anisotropy ($L=$~220~nm, $W=$~50~nm, see Fig.~\ref{fig1}). (Bottom) Temperature dependence of the normalized root mean square UCF amplitude $\delta G^{rms}.L^{3/2}$, showing a universal scaling law $1/T^{3/4}$. Dashed lines define the accuracy for the determination of the prefactor. All results correspond to a perpendicular anisotropy, but open squares (planar anisotropy, $L=$~1200~nm and 600~nm, $W=$~150~nm).

\item UCF measured at base temperature, $T_{el}\approx$~100~nm, for a nanowire with either a perpendicular or a planar anisotropy. (Inset) Crossover in the low-field behavior for a nanowire with a planar anisotropy, showing the effect of domain walls on quantum interferences.

\end{list}

\newpage

\newpage

\newpage

\newpage

\begin{references}

\bibitem{Imry97}
Y.~Imry, \emph{Introduction to Mesoscopic Physics} (Oxford University Press, Oxford, 1997).

\bibitem{Akkermans05}
E.~Akkermans and G.~Montambaux, \emph{Mesoscopic physics of electrons and photons} (EDP Sciences, Paris, 2006).

\bibitem{Benoit88}
A.~Benoit, S.~Washburn, R.A.~Webb, D.~Mailly, and L.~Dumoulin, in \emph{Anderson Localization}, edited by T.~Ando and H.~Fukuyama (Springer, Berlin, 1988). 

\bibitem{Benoit92}
A.~Benoit, D.~Mailly, P.~Perrier, P.~N\'edellec, Superlattices and Microstructures \textbf{11}, 313 (1992). 

\bibitem{Pierre03}
F.~Pierre {\it et al.}, Phys. Rev. B \textbf{68}, 085413 (2003).

\bibitem{Hong95}
K.~Hong and N.~Giordano, Phys. Rev. B \textbf{51}, 9855 (1995).

\bibitem{Tatara97}
G.~Tatara and H.~Fukuyama, Phys. Rev. Lett. \textbf{78}, 3773 (1997).
 
\bibitem{Lyanda98}
Y.~Lyanda-Geller, I.L.~Aleiner, and P.M.~Goldbart, Phys. Rev. Lett. \textbf{81}, 3215 (1998).

\bibitem{Dugaev01}
V.K.~Dugaev, P.~Bruno, and J.~Barn\'as, Phys. Rev. B \textbf{64}, 144423 (2001).

\bibitem{Tatara01}
G.~Tatara and B.~Barbara, Phys. Rev. B \textbf{64}, 172408 (2001). 

\bibitem{Tatara03}
G.~Tatara, H.~Kohno, E.~Bonet, and B.~Barbara, Phys. Rev. B \textbf{69}, 054420 (2003). 

\bibitem{Adam06}
S.~Adam, M.~Kindermann, S.~Rahav, and P.W.~Brouwer, Phys. Rev. B \textbf{73}, 212408 (2006).

\bibitem{Aprili97}
M.~Aprili, J.~Lesueur, L.~Mumoulin, and P.~N\'edellec, Solid State Commun. \textbf{102}, 41 (1997).

\bibitem{Kasai03}
S.~Kasai, E.~Saitoh, and H.~Miyajima, J. Appl. Phys. \textbf{93}, 8427 (2003).

\bibitem{Lee04}
S.~Lee, A.~Trionfi, and D.~Natelson, Phys. Rev. B \textbf{70}, 212407 (2004).
 
\bibitem{Wei06}
Y.G.~Wei, X.Y.~Liu, L.Y.~Zhang, and D.~Davidovic, Phys. Rev. Lett. \textbf{96}, 146803 (2006).

\bibitem{Jaro95}
J.~Jaroszy\'nski {\it et al.}, Phys. Rev. Lett. \textbf{75}, 3170 (1995).

\bibitem{Ohno99}
H.~Ohno, J. Magn. Mag. Mat. \textbf{200}, 110 (1999).

\bibitem{Wagner06}
K.~Wagner {\it et al.}, Phys. Rev. Lett. \textbf{97}, 056803 (2006).

\bibitem{Thevenard06}
L.~Thevenard {\it et al.}, Phys. Rev. B \textbf{73}, 195331 (2006).

\bibitem{note1} A minimum electronic temperature $T_{el}\approx$~100~mK at the cryostat base temperature was infered from the saturation of the longitudinal resistance $R_{long}$ measured on long nanowires to average out the UCF contribution. Besides, $R_{long}$ has an inverse square-root temperature dependence, as expected from Coulomb interactions.
 
\bibitem{Stone87}
P.A.~Lee, A.D.~Stone, and H.~Fukuyama, Phys. Rev. B \textbf{35}, 1039 (1987).

\end{references}
\end{document}